\documentclass[showpacs,superscriptaddress,amsmath,amssymb,aps,preprint]{revtex4}
\usepackage{graphicx}
\usepackage{bm}
\usepackage[utf8]{inputenc}
\usepackage{lmodern}
\begin{document}
\title{Wave dispersion derived from the square-root Klein-Gordon-Poisson system} 
\author{F. Haas}
\affiliation{Departamento de F{\'i}sica, Universidade Federal do Paran\'a, 81531-990, Curitiba, Paran\'a, Brazil}
\begin{abstract}
Recently there has been great interest around quantum relativistic models 
for plasmas. In particular striking advances have been obtained by means 
of the Klein-Gordon-Maxwell system, which provides a first order approach 
to the relativistic regimes of quantum plasmas. It is a reliable method 
as long as the plasma spin dynamics is not a fundamental aspect, to be 
addressed using more refined (and heavier) models involving the Pauli-Schr\"odinger 
or Dirac equations. In this work a further simplification is considered, 
tracing back to the early days of relativistic quantum theory. 
Namely, we revisit the square-root Klein-Gordon-Poisson system,
where the positive branch of the relativistic energy-momentum relation is 
mapped to a quantum wave equation. The associated linear wave propagation is 
analyzed and compared to the results in the literature. We determine  
physical parameters where the simultaneous quantum and relativistic effects 
can be noticeable in weakly coupled electrostatic plasmas. 
\end{abstract}
\pacs{03.65.Pm, 52.25.Dg, 52.27.Ny, 52.35.-ge}
\maketitle
\section{Introduction}
Recently there has been much interest in the analysis of relativistic effects in quantum plasmas, 
due to potential applications in the realm of multi-petawatt lasers and dense astrophysical objects
like white dwarfs and neutron stars 
Promising new models have appeared,
like in weakly relativistic expansions including spin dynamic effects \cite{Asenjo}, hydrodynamic 
equations derived from covariant quantum kinetics \cite{Zhu}, collective Klein-Gordon-Maxwell models \cite{Eliasson},
quantum relativistic multi-stream approaches \cite{Haas1, Haas2}, quantum kinetic theories derived from 
the Klein-Gordon-Maxwell system \cite{Tito} as well as the quantum plasmadynamics approach \cite{Melrose}. There 
is also a recent study on Landau damping in relativistic quantum kinetics of degenerate plasma \cite{Zhu2}.

In the present communication we revisit the semi-relativistic quantum kinetic model introduced in \cite{Markowich}, which originate 
from the simplest possible quantum relativistic theory, based on the square-root Klein-Gordon equation \cite{Bjorken}. 
Here we explore the implications of such a model, with respect to plasma behavior i.e. wave propagation. We are aware of the 
intrinsic limitations of the square-root Klein-Gordon approach, to be detailed in the continuation. However, since it is 
the simplest approach, it is worth to be discussed in some detail, as a benchmark for more sophisticated analysis. The 
resulting robust modeling, not too expensive analytically or numerically, can then produce useful insights to complex 
quantum relativistic phenomena. Also, the square-root Klein-Gordon equation has had applications e.g. in the study of 
subbarrier relativistic effects in heavy nucleons penetration coefficients \cite{Anchishkin}.

This work is organized as follows. In Section II we rederive the quantum kinetic theory associated to the square-root Klein-Gordon-Poisson system, as already done in \cite{Markowich}. A distinctive feature to this reference is that here the 
evolution equation satisfied by the non-covariant Wigner function is set into a manifestly Schr\"odinger-like form. In Section III the dispersion relation for linear electrostatic waves is obtained. A new quantum modified relativistic factor is then identified. An application is devised for a mono-energetic beam equilibrium. Physically relevant parameters are found, where the joint quantum 
and relativistic effects significantly alter the wave propagation. In addition the underlying validity conditions and limitations of 
the model are considered. Section IV has the concluding remarks. 

\section{\label{sec2}Square-root Klein-Gordon-Poisson system and Wigner transform}
Our starting point is the Hamiltonian 
\begin{equation}
\label{e1}
H = m\,c^2 \left(1 + \frac{p^2}{m^2 c^2}\right)^{1/2} - e\,\phi \,,
\end{equation}
where $-e$ and $m$ are resp. the electron charge and rest mass, $c$ is the speed of light, ${\bf p}$ is the electron momentum (with $p = |{\bf p}|$) and $\phi = \phi({\bf r},t)$ is the electrostatic potential at position ${\bf r}$ and time $t$. Equation (\ref{e1}) is consistent with the energy-momentum relation
\begin{equation}
\label{e2}
(H + e\phi)^2 = p^2 c^2 + m^2 c^4 \,,
\end{equation}
disregarding the negative energy solutions associated with the 
positron sector of Hilbert space. Assuming $i\hbar\partial\psi/\partial t = H\psi$ and applying the quantization rules ${\bf p} \rightarrow - i \hbar \nabla$ and $E \rightarrow i \hbar \partial/\partial t$ one obtains 
\begin{equation}
\label{e3}
i\,\hbar\frac{\partial\psi}{\partial t} = \left[m\, c^2 \left(1 - \frac{\hbar^2 \nabla^2}{m^2 c^2}\right)^{1/2} - e\phi\right]\psi \,,
\end{equation}
where $\hbar$ is Planck's constant divided by $2\pi$ and $\psi = \psi({\bf r},t)$ is the wave function. 

The square-root Klein-Gordon equation (\ref{e3}) was proposed in the early days of relativistic quantum mechanics \cite{Bjorken}, but soon abandoned in favor of the Klein-Gordon and Dirac equations, for particles with spin zero and one-half, respectively. The main drawback with Eq. (\ref{e3}) is that time and space appear in a non-symmetric way in it, spoiling manifest covariance. Moreover, the square-root operator poses analytic difficulties, since it needs to be implemented using Fourier-transformed variables along with an infinite power series expansion. In other words, one is obliged to deal with a nonlocal operator. The introduction of electromagnetic fields in a relativistic invariant manner is hence apparently an unfeasible task within the framework of the the square-root Klein-Gordon equation. Indeed, using the minimal coupling scheme one would deal with infinite powers of the quantity ${\bf p} + e{\bf A}$, where ${\bf A} = {\bf A}({\bf r},t)$ is the vector potential. This imply a troublesome problem due to non-commutativeness of momentum and position, although in principle solvable up to arbitrary desired order. In addition, one can still wonder about alternative schemes to interpret the square-root operator in the presence of magnetic fields \cite{Gill}, at the price of some lost of simplicity. 

Evidently, Eq. (\ref{e1}) does not take into account retardation effects nor space-time curvature, besides magnetic fields or spin. Hence we are dealing with a kind of semi-relativistic model. Not surprisingly, similar semi-relativistic models are popular in the treatment of boson stars \cite{Michelangeli}, where N-body bosons systems interact gravitationally, without the need of general relativity nor magnetic forces to be included. The same modeling but also for fermionic systems was treated in \cite{Lieb, Frohlich}, in connection with the analysis of the gravitational collapse of white dwarfs and the related Chandrasekhar limit.  Using the same reasoning, the Hamiltonian (\ref{e1}) can be seriously considered as the starting point for electrostatic relativistic quantum plasmas - hence excluding any magnetic interaction. Moreover, we will observe that the practical implementation of the square-root operator in the context of quantum kinetic theory poses no extra difficulties, in comparison with the already existing nonlocal potential term in the quantum Vlasov equation. Since quantum plasma theory is usually more concerned with kinetic formulations than the wave equations themselves, in this respect one sees no profound reason to discard the semi-relativistic formulation. Finally, note that even within the electron sector of Hilbert space and in the electrostatic approximation one has an extra $\propto \nabla\cdot{\bf E}$ Darwin term in the weakly relativistic approximation to the Dirac equation \cite{Zawadzki}. This term, related to {\it Zitterbewegung} and the associated enlarged expected value of the quadratic displacement, can be shown to be of higher order and will be ignored in what follows.  

After all, the motivation in considering the square-root Klein-Gordon model is in its simplicity. Indeed, in the case of electrostatic quantum plasma, it offer the simplest conceivable method to incorporate relativistic effects. In particular, the corresponding kinetic theory can be constructed from a non-covariant Wigner function, in contrast to the Klein-Gordon case which deserves 
a Wigner transform in both space and time \cite{Tito}. This follows since Eq. (\ref{e3}) clearly breaks down the symmetry between space and time variables. From one side, this is a further limitation, since any relativistic theory should be preferably covariant. From the other side, it is conceptually useful to have a relativistic model which can be directly comparable to the non-relativistic version (the Wigner-Poisson system) in quantum plasmas. In this way we have a benchmark model, allowing a zeroth order understanding of relativistic phenomena in quantum plasma. In \cite{Tito} the Klein-Gordon-Maxwell system was referred to as a ``first order solution for a very difficult problem". In this sense the square-root version may be named a ``zeroth order solution".

Equation Eq. (\ref{e3}) is coupled with Poisson's equation,
\begin{equation}
\label{e4}
\nabla^2 \phi = \frac{e}{\varepsilon_0} (|\psi|^2 - n_0) \,,
\end{equation}
where $\varepsilon_0$ is the vacuum permittivity. We suppose a jellium plasma with uniform background ionic density $n_0$ to ensure global charge neutrality. The system (\ref{e3})--(\ref{e4}) can be entitled the square-root Klein-Gordon-Poisson system. The normalization $\int d{\bf r} |\psi|^2 = N =$ number of particles of the system is used. 

It should be noted that a more general version can be encountered in the literature, allowing for mixed states \cite{Markowich}. In this case one has a countable set of square-root Klein-Gordon equations for the quantum statistical ensemble, with self-consistent scalar field mediated by Poisson's equation. Our attempt here is to address the plasma aspects of the model, specially the wave 
dispersion analysis. For these issues, a pure state ensemble (a single wave function) $\psi$ is sufficient, at least to begin with. 

It is useful to write our model in terms of a kinetic theory, so as to provide a relativistic version of the Wigner-Poisson system. Following \cite{Markowich}, we Wigner-transform Eq. (\ref{e3}) by means of the non-covariant Wigner function \cite{Wigner}, 
\begin{equation}
\label{e5}
f({\bf p},{\bf r}) = \frac{1}{(2\pi\hbar)^3} \int d{\bf r}'\, \psi^{*}\left({\bf r} + \frac{{\bf r}'}{2}\right) \exp\left(\frac{i\,\,{\bf p}\cdot{\bf r}'}{\hbar}\right) \psi\left({\bf r} - \frac{{\bf r}'}{2}\right) \,.
\end{equation}
In Eq. (\ref{e5}) and henceforth, for the sake of notation the explicit time-dependence is omitted unless necessary.

For completeness we rederive the evolution equation satisfied by the Wigner function \cite{Markowich}. Consider the Fourier transform in space 
along with its inverse, 
\begin{equation}
\label{e8}
\hat{\psi}({\bf k}) = \frac{1}{(2\pi)^{3/2}}\int d{\bf r}\, \psi({\bf r})\, e^{- i\,{\bf k}\cdot{\bf r}} \,, \quad \psi({\bf r}) = \frac{1}{(2\pi)^{3/2}}\int d{\bf k}\, \hat{\psi}({\bf k})\, e^{i\,{\bf k}\cdot{\bf r}} \,,
\end{equation}
with associated wave vector ${\bf k}$. One find 
\begin{equation}
\label{ew}
f({\bf p},{\bf r}) = \frac{1}{(2\pi\hbar)^3}\int d{\bf k}\,\hat{\psi}^{*}\left(\frac{\bf p}{\hbar}+\frac{\bf k}{2}\right)\,e^{-i\,{\bf k}\cdot{\bf r}}\hat{\psi}\left(\frac{\bf p}{\hbar}-\frac{\bf k}{2}\right) \,.
\end{equation}

Momentarily only the kinetic energy term will be considered since the scalar potential contribution to $\partial f/\partial t$ 
is given in terms of a well-known pseudo-differential operator \cite{Haas3}. 
In this context one get 
\begin{equation}
\label{e9}
i\,\hbar\frac{\partial\hat{\psi}}{\partial t} = m\, c^2 \left(1 + \frac{\hbar^2 k^2}{m^2 c^2}\right)^{1/2}\hat{\psi} \,, \quad i\,\hbar\frac{\partial\hat{\psi}^{*}}{\partial t} = - m\, c^2 \left(1 + \frac{\hbar^2 k^2}{m^2 c^2}\right)^{1/2}\hat{\psi}^{*} \,, \quad k = |{\bf k}| \,.
\end{equation}
Notice that Eq. (\ref{e9}) provides a way to interpret the square-root operator, upon its action on dual space. The inverse Fourier transform is an integral - in this sense the square-root operator act in a nonlocal way. 

Equations (\ref{ew}) and (\ref{e9}) yield 
\begin{eqnarray}
\nonumber
\frac{\partial f}{\partial t} &=& - \frac{m\, c^2}{i\,\hbar (2\pi\hbar)^3} \int d{\bf k}\, \hat{\psi}^{*}\left(\frac{\bf p}{\hbar}+\frac{\bf k}{2}\right)\,e^{- i\,{\bf k}\cdot{\bf r}} \hat{\psi}\left(\frac{\bf p}{\hbar}-\frac{\bf k}{2}\right) \times \\
\label{e10}
&\times& \left[\left(1 + \frac{1}{m^2 c^2}\Bigl({\bf p} + \frac{\hbar\,{\bf k}}{2}\Bigr)^2\right)^{1/2} - \left(1 + \frac{1}{m^2 c^2}\Bigl({\bf p} - \frac{\hbar\,{\bf k}}{2}\Bigr)^2\right)^{1/2}\right] \,.
\end{eqnarray}

To proceed consider the identity
\begin{equation}
\label{e11}
\hat{\psi}^{*}\left(\frac{\bf p}{\hbar}+\frac{\bf k}{2}\right) \hat{\psi}\left(\frac{\bf p}{\hbar}-\frac{\bf k}{2}\right) = \hbar^3 \int d{\bf r} f({\bf p},{\bf r})\,e^{i\,{\bf k}\cdot{\bf r}}  \,,
\end{equation}
following from Eq. (\ref{ew}). From Eqs. (\ref{e10}) and (\ref{e11}) and restoring the scalar potential term we finally obtain the Schr\"odinger-like 
equation 
\begin{equation}
\label{schro}
i\hbar\frac{\partial f}{\partial t} = {\cal H}[f] \,,
\end{equation}
where ${\cal H}[f]$ is a functional defined by 
\begin{eqnarray}
\nonumber
{\cal H}[f] &=& m\,c^2 \int\frac{d{\bf p}'\,d{\bf r}'}{(2\pi\hbar)^3}\, f({\bf p},{\bf r}')\,\exp\left[\frac{i\,\,{\bf p}'\cdot({\bf r}-{\bf r}')}{\hbar}\right] \times \\  \label{e12}
&\times& \left[\left(1 + \frac{1}{m^2 c^2}\Bigl({\bf p} + \frac{{\bf p}'}{2}\Bigr)^2\right)^{1/2} - \left(1 + \frac{1}{m^2 c^2}\Bigl({\bf p} - \frac{{\bf p}'}{2}\Bigr)^2\right)^{1/2}\right]  \\
&+& e \int\frac{d{\bf p}'\,d{\bf r}'}{(2\pi\hbar)^3}\, f({\bf p}',{\bf r})\,\exp\left[\frac{i\,\,({\bf p}-{\bf p}')\cdot{\bf r}'}{\hbar}\right] \left[\phi\left({\bf r} + \frac{{\bf r}'}{2}\right) - \phi\left({\bf r} - \frac{{\bf r}'}{2}\right)\right] \,. \nonumber
\end{eqnarray}
A very detailed derivation of the non-local potential energy term can be found e.g. in \cite{Haas3}. The distinctive feature of the relativistic extension is the extra non-locality also appearing in the kinetic energy term. Coupling with Poisson's equation in the form
\begin{equation}
\label{e13}
\nabla^2 \phi = \frac{e}{\varepsilon_0}\left(\int d{\bf p} f({\bf p},{\bf r}) - n_0\right)
\end{equation}
gives a quantum relativistic generalization of the Vlasov-Poisson system \cite{Markowich}.

In the formal non-quantum ($\hbar \rightarrow 0$) limit it is easy to obtain
\begin{equation}
\label{e14}
\frac{\partial f}{\partial t} + \frac{{\bf p}\cdot\nabla f}{\gamma m} + e\nabla\phi\cdot\frac{\partial f}{\partial{\bf p}} = 0 \,,
\end{equation}
which is the relativistic Vlasov equation, where $\gamma = \left[1+p^2/(m^2 c^2)\right]^{1/2}$. Here, evidently ``$\hbar \rightarrow 0$" is just a shorthand notation,  denoting the appropriate limit when quantum diffraction is negligible. On the same footing, the 
formal non-relativistic ($1/c \rightarrow 0$) limit yield 
\begin{equation}
\label{e15}
\frac{\partial f}{\partial t} + \frac{{\bf p}\cdot\nabla f}{m} - \frac{e}{i\hbar}\int\frac{d{\bf p}'\,d{\bf r}'}{(2\pi\hbar)^3}f({\bf p}',{\bf r})\,\exp\left[\frac{i\,\,({\bf p}-{\bf p}')\cdot{\bf r}'}{\hbar}\right]\left[\phi\left({\bf r} + \frac{{\bf r}'}{2}\right) - \phi\left({\bf r} - \frac{{\bf r}'}{2}\right)\right] = 0 \,,
\end{equation}
which is the well-known non-relativistic quantum Vlasov equation. 

Observe that starting with the Klein-Gordon equation one arrive at an equation whose non-relativistic limit is {\it not} the usual quantum Vlasov equation (see Eq. (11) of \cite{Tito}). The reason is that in a covariant formalism the time is also Wigner-transformed. It is useful to have an alternative tool allowing a more straightforward comparison with the well-known non-relativistic theory. Further, one may argue that the extra non-local contribution in Eq. (\ref{e12}) poses no fundamental difficulties, since it can handled basically in the same way as the potential energy term. It is interesting to note that the relativistic quantum Vlasov equation satisfied by the non-covariant Wigner equation shows a perfect symmetry between momentum and coordinate variables, in the same sense as ${\cal H}[f]$ has both the kinetic and potential terms expressed as integrals. 

\section{\label{sec3}Wave dispersion}
\begin{figure}
\includegraphics[height=.3\textheight]{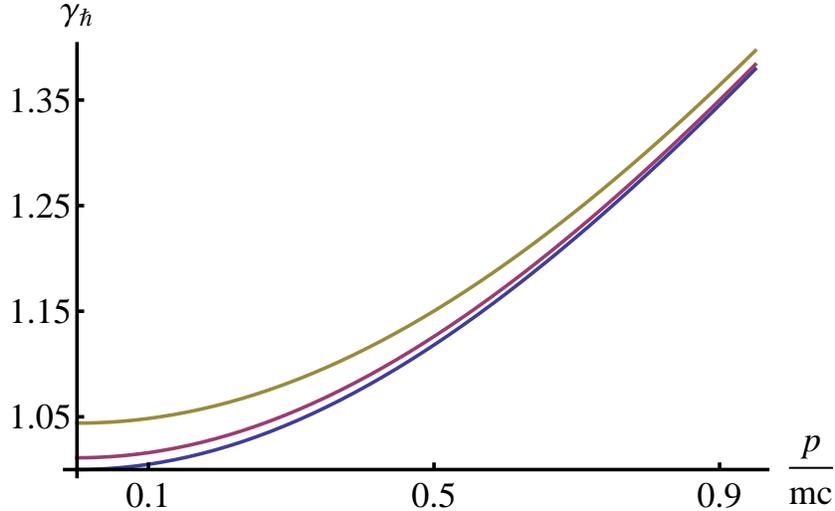}
\caption{Quantum modified relativistic factor from Eq. (\ref{e18}), assuming ${\bf k}\parallel{\bf p}$, with $\hbar k = 0$ (bottom curve), $\hbar k = 0.3 m c$ (mid curve) and $\hbar k = 0.6 m c$ (top curve)} 
\label{fig1}
\end{figure}

The main concern of the present communication is about the linear wave propagation implied by the system (\ref{schro}) and (\ref{e13}). 
Linearizing around the equilibrium $f = f_{0}({\bf p}), \phi = 0$, assuming plane wave perturbations proportional to $\exp[i({\bf k}\cdot{\bf r} - \omega t)]$ and 
following the standard procedure one find the dielectric function
\begin{equation}
\label{e17}
\varepsilon(\omega,{\bf k}) = 1 - \frac{m\omega_{p}^2}{n_0\hbar\,k^2}\int\,d{\bf p}\,\frac{f_{0}({\bf p} + \hbar{\bf k}/2) - f_{0}({\bf p}-\hbar{\bf k}/2)}{{\bf k}\cdot{\bf p}/(\gamma_{\hbar}m) - \omega} = 0 \,,
\end{equation}
where $\omega_p = [n_0 e^2/(m\varepsilon_0)]^{1/2}$ is the plasma frequency and where
\begin{equation}
\gamma_\hbar = \frac{\hbar\,{\bf k}\cdot{\bf p}}{m^2 c^2}\left[\left(1 + \frac{1}{m^2 c^2}\Bigl({\bf p}+\frac{\hbar{\bf k}}{2}\Bigr)^2\right)^{1/2}  - 
\left(1 + \frac{1}{m^2 c^2}\Bigl({\bf p}-\frac{\hbar{\bf k}}{2}\Bigr)^2\right)^{1/2}\right]^{-1}  \label{e18} 
\end{equation}
is a quantum modified relativistic factor (which is different from the one in \cite{Markowich}). The semi-quantum and semi-relativistic expansions are, 
respectively,  
\begin{equation}
\gamma_\hbar = \gamma + \frac{\hbar^2}{8\, m^2 c^2 \gamma^3}\left(k^2 + \frac{|{\bf k}\times{\bf p}|^2}{m^2 c^2}\right) + O(\hbar^4) \,,
\end{equation}
and 
\begin{equation}
\gamma_\hbar = 1 + \frac{p^2 + \hbar^2 k^2/4}{2\, m^2 c^2} + O(1/c^4) \,.
\end{equation}
We see that the contribution of the quantum recoil is for a larger $\gamma_\hbar$ and 
consequently in favor of a larger effective mass, as confirmed in Figure \ref{fig1}.

Expanding up to the leading relativistic correction the result is 
\begin{eqnarray}
\varepsilon(\omega,{\bf k}) = 1 &-& \frac{m\omega_{p}^2}{n_0\hbar\,k^2}\int\,d{\bf p}\,\left[\frac{f_{0}({\bf p} + \hbar{\bf k}/2) - f_{0}({\bf p}-\hbar{\bf k}/2)}{{\bf k}\cdot{\bf p}/m - \omega}\right] \times \nonumber \\
\label{e19}
&\times& \left[1 + \frac{{\bf k}\cdot{\bf p}\,(p^2 + \hbar^2 k^2/4)}{2\, m^3 c^2 ({\bf k}\cdot{\bf p}/m - \omega)}\right]  \,.
\end{eqnarray}
In Eq. (\ref{e19}) the $O(1/c^4)$ contribution is disregarded. In practice, the weakly relativistic approximation may be quite useful for 
calculations since the resonances in Eq. (\ref{e17}) are not simple poles. The same apply to the purely relativistic and non-quantum case too, but the quantum situation is worse due to the finite-difference and nonlinear form of $\gamma_\hbar$.  

In this work we limit ourselves to a further approximation, namely the simultaneous weakly quantum and weakly relativistic version of the dielectric function. This amounts to keeping only the leading quantum recoil correction in Eq. (\ref{e19}) (also disregarding the joint quantum and relativistic perturbations which are $O(\hbar^2/c^2)$). Assuming decaying boundary conditions and integrating by parts the result is
\begin{eqnarray}
\label{sqr}
\varepsilon(\omega,{\bf k})\!\!\!\!&=& 1 - \frac{\omega_{p}^2}{n_0}\int\frac{d{\bf p} f_{0}({\bf p})}{(\omega - {\bf k}\cdot{\bf p}/m)^2}  
- \frac{\hbar^2 k^4 \omega_{p}^2}{4 m^2 n_0}\int\frac{d{\bf p} f_{0}({\bf p})}{(\omega - {\bf k}\cdot{\bf p}/m)^4}  
\\
&-&\!\!\frac{\omega_{p}^2}{2n_0 m^2 c^2 k^2}\int\frac{d{\bf p} f_{0}({\bf p})}{(\omega - {\bf k}\cdot{\bf p}/m)^3}\left[2({\bf k}\cdot{\bf p})^2\left(\frac{{\bf k}\cdot{\bf p}}{m}-\omega\right) - p^2 k^2 \left(\frac{{\bf k}\cdot{\bf p}}{m}+\omega\right)\right] \,. \nonumber 
\end{eqnarray}

Consider the simplest possible case, a mono-energetic beam,
\begin{equation}
f_{0}({\bf p}) = n_0 \,\delta({\bf p} - {\bf p}_0) \,,
\end{equation}
where for definiteness we take ${\bf p}_0 \parallel {\bf k}$. Observe that this a quantum-mechanically acceptable equilibrium, associated to infinite precision in momentum space and complete delocalization in position. More exactly it corresponds to the plane wave $\psi = \sqrt{n_0}\exp(i{\bf p}_{0}\cdot{\bf r}/\hbar)$. 

Inserting in Eq. (\ref{sqr}) and solving recursively the dispersion relation up to $O(\hbar^2)$ and $O(1/c^2)$ results in
\begin{equation}
\label{w}
\omega \simeq \omega_{p}\left(1 - \frac{3p_{0}^2}{4m^2 c^2}\right) + \frac{p_0 k}{m}\left(1 - \frac{p_{0}^2}{2m^2 c^2}\right) + \frac{\hbar^2 k^4}{8m^2\omega_p} \,,
\end{equation}
which can be checked to be consistent either with the exact non-relativistic
\begin{equation}
\left(\omega - \frac{p_0 k}{m}\right)^2 = \omega_{p}^2 + \frac{\hbar^2 k^4}{4m^2} \quad\quad (1/c \rightarrow 0) 
\end{equation}
or exact non-quantum
\begin{equation}
\left(\omega - \frac{p_0 k}{\gamma m}\right)^2 = \frac{\omega_{p}^2}{\gamma^3} \quad\quad (\hbar \rightarrow 0) 
\end{equation}
dispersion relations. 

To proceed, we consider the case where the Doppler correction is much larger than the plasma frequency,
\begin{equation}
\label{d}
\frac{p_0 k}{m} \gg \omega_p \,.
\end{equation}
The motivation for this assumption is to pay attention to the cases where the relativistic correction is significant. Hence Eq. (\ref{w}) reduces to 
\begin{equation}
\label{ww}
\omega \simeq \frac{p_0 k}{m}\left(1 - \frac{p_{0}^2}{2m^2 c^2}\right) + \frac{\hbar^2 k^4}{8m^2\omega_p} \,,
\end{equation}
From this simple expression it is evident that quantum diffraction act in the opposite direction of the relativistic effects, which is in 
accordance with recent findings on quantum corrected wake field acceleration \cite{Zhu}. Moreover for large enough wave numbers such that 
\begin{equation}
\label{k}
\frac{\hbar^3 k^3}{p_{0}^3} = \frac{4\,\hbar \omega_p}{m c^2} 
\end{equation}
the quantum and relativistic corrections cancel out. Let us examine the implications of this result in concrete examples. First of all we 
explicitly write the assumptions leading to Eq. (\ref{k}): (a) weakly relativistic and quantum effects; (b) large Doppler shift, Eq. (\ref{d}); (c) cold beam and weak coupling approximations; (d) no QED effects, or $\hbar\omega \ll 2mc^2$. One further limitation is due to the neglect of the return current and the associated magnetic field generation. This approximation can be justified for a sufficiently small beam cross section, as discussed in detail in \cite{Bludman}.

Condition (c) is equivalent to 
\begin{equation}
\label{t}
K \gg {\rm max}\{\kappa_B T, E_F\} \gg U \,,
\end{equation}
where $\kappa_B$ is the Boltzmann constant, $T$ is the bulk temperature and 
\begin{equation}
K = mc^2 (\gamma -1) \,, \quad E_F = \left(m^2 c^4 + \hbar^2 c^2 (3\pi^2 n_0)^{2/3}\right)^{1/2} - m c^2 \,, \quad U = \frac{e^2 n_{0}^{1/3}}{4\pi\varepsilon_0}
\end{equation}
are resp. the beam kinetic energy, the Fermi energy (both in their relativistic form) and a measure of the electrostatic energy. The inequalities (\ref{t}) allow us 
to neglect both thermal and collisional contributions. From now on assume a beam with speed $v_0 = 0,4c$, which is weakly relativistic since then $p_{0}^2/(2m^2c^2) = 0,1$ (or $10\%$ relativistic correction in the frequency), where $p_0 = \gamma m v_0$. Equivalently the beam's kinetic energy is set as $K = 46,7 \,keV$.

We can consider first a non-degenerate plasma with $n_0 = 10^{24} m^{-3}$. In this case Eq. (\ref{k}) gives $k = 7,5 \times 10^9 m^{-1}$ and a wavelength $\lambda = 2\pi/k = 8,4 \times 10^{-10} m$. Such a plasma is indeed non-degenerate, because Eq. (\ref{t}) imply $46,7 \,keV \gg T \gg 0,1 \,eV$, which always exceeds by far the Fermi energy $E_F = 3,7\, meV$. Moreover in such conditions the large Doppler and negligible QED assumptions are fairly well satisfied. From Eq. (\ref{ww}) one find $\omega = 9,8 \times 10^{17}\,Hz$.

Secondly, consider a large density $n_0 = 10^{32} m^{-3}$, which can be attained in intense laser compression experiments. From (\ref{k}) one get $k = 1,6 \times 10^{11} m^{-1}$ as the wavenumber for which quantum diffraction equals the relativistic correction, or $\lambda = 3,9 \times 10^{-11} m$ (in the hard X ray regime). For temperatures smaller than $E_F = 787,0 \,eV$ such a plasma is degenerate. Moreover one find $E_{F}/K = 0,02$ and $U/E_{F} = 0,09$ , in agreement with the cold beam and weak coupling hypothesis. Finally, the large Doppler shift and negligible QED effects are also fairly well verified. From Eq. (\ref{ww}) one find $\omega = 2,1 \times 10^{19}\,Hz$.

\section{\label{sec6}Conclusion}
In this work we analyzed the electrostatic waves following a  
semi-relativistic Wigner-Poisson system given by the Schr\"odinger-like equation (\ref{schro}) and Poisson's equation (\ref{e13}). A linear 
dispersion relation containing a quantum modified relativistic factor was deduced. Afterwards the case of a mono-energetic beam was discussed and 
sample physical parameters allowing for jointly significant relativistic and quantum effects were identified. From the last Section one conclude that 
the quantum diffraction effects are more easily measurable in non-degenerate plasmas, since they correspond to wavelengths on the nanoscale in this case. For larger densities we found that the relativistic correction is dominant except for extremely small spatial scales. 

In spite of the overall simplification which limit the applicability of the square-root Klein-Gordon approach, it is nevertheless an useful method for 
direct comparison with the known results for traditional plasma. A further limitation which is worth to comment is that the integro-differential equation (\ref{schro}) for the Wigner function makes it hardly accessible to nonlinear analysis, except perhaps by means of numerical simulation. This is a common disadvantage in any kinetic theory.

We expect that the parameter analysis of the last Section can inspire similar approaches to 
serious investigation of the relevance (or irrelevance) of quantum diffraction effects in specific plasma problems. The study of more structured equilibria than a single beam is also recommendable. 

\vspace{1cm}
{\bf Ack\-now\-ledgments}\\

This work was supported by Conselho Nacional de Desenvolvimento Cien\-t\'{\i}\-fi\-co e Tec\-no\-l\'o\-gi\-co (CNPq).


\end{document}